# Artificial Intelligence Training in Media: Addressing Technical and Ethical Challenges for Journalists and Media Professionals

Barbara Sarrionandia[1], Simón Peña-Fernández[2*], Jesús Ángel Pérez Dasilva and Ainara Larrondo-Ureta[4]

1. University of the Basque Country (UPV/EHU) (ROR: 000xsnr85)
   bsarrionaindia001@ikasle.ehu.eus / https://orcid.org/0009-0006-0053-3730
2. University of the Basque Country (UPV/EHU) (ROR: 000xsnr85)
   simon.pena@ehu.eus / https://orcid.org/0000-0003-2080-3241
3. University of the Basque Country (UPV/EHU) (ROR: 000xsnr85)
   jesusangel.perez@ehu.eus / https://orcid.org/0000-0002-3383-4859
4. University of the Basque Country (UPV/EHU) (ROR: 000xsnr85)
   ainara.larrondo@ehu.eus / https://orcid.org/0000-0003-3303-4330

* Corresponding author

**Abstract**: The rise of Artificial Intelligence (AI) is presenting both technical and ethical challenges for media organisations, creating an urgent need for professional training. This study explores how media professionals in the Basque Country are equipping themselves to face these challenges. Using a mixed-method approach, it combines a survey of 504 active professionals with in-depth interviews with six innovation leaders from major regional media outlets. The findings reveal that only 14.1% of professionals have undergone AI training, mostly through self-learning. Larger, internationally focused companies are more proactive in providing training, while local and traditional media organisations show significant gaps. Technical and managerial roles are leading the way in adopting AI, whereas newsroom staff are notably behind. The study highlights the pressing need to enhance AI training, with a particular focus on ethical and technical aspects, both through in-house programmes and formal education pathways.

## 1. Introduction

The adoption of AI in newsrooms and media organisations has introduced a range of complex challenges for journalists and media professionals (Beckett et al., 2023; Coeckelbergh, 2020). As news production and distribution evolve, AI offers opportunities for efficiency and innovation, but it also brings risks and vulnerabilities (Brennen et al., 2018).

Much like previous technological advancements in the digitalisation of media, AI is a tool that requires careful consideration of its ethical implications. Challenges such as reinforcing biases through technology (Leiser, 2022), increasing misinformation and manipulation (García-Marín and Salvat-Martinrey, 2022), and safeguarding privacy, can all be mitigated with robust media literacy and targeted training. These measures empower journalists to not only use AI tools effectively but also navigate the ethical dilemmas they present (Aissani et al., 2023; Baldessar and Zandomênico, 2024; Barceló-Ugarte et al., 2021; Helberger and Diakopoulos, 2023; Ufarte Ruiz et al., 2021).

AI, with its continuous evolution, holds the potential to transform our lives, work, and relationships. However, it also raises serious ethical and social justice concerns, particularly regarding the widening of existing disparities (e.g., race and gender) (Brundage et al., 2018) and the growing sophistication and dissemination of misinformation, especially through advanced deepfake creation (Phelan, 2022; Sohrawardi et al., 2024).

Nonetheless, AI is not solely a cause for concern among journalism professionals (Peña-Fernández et al., 2023a). In its more constructive application, it can serve as a powerful tool to combat informational disorders and malicious content (Bontridder and Poullet, 2021; Manfredi-Sánchez and Ufarte-Ruiz, 2020; Rubin, 2022). This technology has the ability to match the speed and sophistication of digital falsehoods, reducing the effort and time required by fact-checking professionals and enhancing their capacity to respond effectively to misinformation.

For AI to strengthen journalism's role in supporting democracy (Lin and Lewis, 2022), it is vital to establish adequate regulations. These should include clear guidelines on authorship (Hofeditz et al., 2021; Díaz-Noci et al., 2024) and enforce transparency (Larsson and Heintz, 2020). AI's integration into media does not only impact editorial workflows but also necessitates significant shifts in the skills and competencies required for journalists (Danzon-Chambaud and Cornia, 2023; Demmar and Neff, 2023; Gómez-Diago, 2022; Lopezosa et al., 2023; Noain Sánchez, 2022; Ufarte-Ruiz et al., 2020).

## 2. State of the art

The prominence of AI, heightened since late 2022 with the emergence of generative AI and tools like ChatGPT, has spurred significant innovation across various fields, including journalism (Bakke and Barland, 2022; García-Orosa et al., 2023; Tejedor and Vila, 2021). Although AI is not a novel concept in media (Manfredi-Sánchez and Ufarte-Ruiz, 2020), efforts to automate news production by major outlets and agencies have provided examples over the past two decades. For instance, The Big Ten Network has used automated systems for sports reporting since 2007, while Quakebot at The Los Angeles Times has been offering real-time earthquake updates since 2014. Agencies such as Reuters and Associated Press have also automated parts of their services (Danzon-Chambaud, 2021; Sánchez-García et al., 2023).

A revolutionary shift has been the democratisation of generative AI tools, with ChatGPT as a leading example. Numerous companies have introduced similar platforms (Motlagh et al., 2023; Singh et al., 2023). The societal adoption of these technologies has been remarkably swift, with ChatGPT reaching 100 million users within just 2 months, outpacing TikTok (9 months) and Instagram (two and a half years) (Milmo, 2023).

The emergence of such tools can be seen as the latest phase in media's digital transformation, redefining its nature within the framework of the Fourth Industrial Revolution (Micó et al., 2022;

Dhiman, 2023). This transformation is expected to have a widespread impact across all areas of communication, from news production processes (Sánchez-García et al., 2023) to adjustments in normative values (Peña-Fernández et al., 2023b).

AI is reshaping journalism in multiple ways. By automating content production and reducing reporters' workload, it fosters innovation and professional specialisation, allowing journalists to focus on more cognitive and creative tasks (Wu et al., 2019). Generative AI can also enable coverage of previously unprofitable topics (Atasoy et al., 2021) and deliver more personalised content (Hermann, 2022).

Integrating AI into newsrooms seems inevitable, as this technology has become essential for the evolution of both media and journalism itself (Terol, 2023; García-Orosa et al., 2023; Gutiérrez-Caneda et al., 2023).

For professionals, concerns about AI extend beyond industrial or sectoral issues such as changes in production processes, potential job losses, or shifts in required skill sets. There is considerable uncertainty about its impact on public opinion and democratic societies.

Since its release in November 2022, OpenAI's platform has raised concerns regarding the spread of misinformation (Aydın and Karaarslan, 2023; Opdahl et al., 2023). Although the March 2023 release of ChatGPT-4 showed improvements in source citation, misinformation remains a persistent issue (Gutiérrez-Caneda et al., 2023).

During this period, discourse among media professionals has increasingly focused on the potential negative consequences of AI in journalism (Beckett et al., 2023). Concerns include the potential for AI reliance to degrade journalistic quality (Calvo-Rubio and Rojas-Torrijos, 2024), reinforce inherent biases (Cloudy et al., 2023; Leiser, 2022), and encourage unethical practices such as content farms and plagiarism (Palacios Tapia, 2023; Subiela-Hernández and Vizcaíno-Laorga, 2023).

Despite these risks, the primary concern among professionals since the advent of generative AI remains its role in spreading false information, fostering misinformation, and exacerbating polarisation (Berrocal-Gonzalo et al., 2023; Peña-Fernández et al., 2023a).

In response to the growing social and political concerns about misinformation, governments and institutions have intensified efforts to mitigate its impact. Recent measures include the development of tools like fact-checking algorithms, detection systems (e.g., random forests), bots and chatbots, and monitoring platforms, all of which heavily rely on AI (Arias Jiménez et al., 2022; Garriga et al., 2024; Moreno Espinosa et al., 2024; Alonso González and Sánchez Gonzales, 2024). These initiatives combine information verification with media literacy programmes, ongoing training, and professional development for journalists.

Given the technical and ethical complexities involved, training for current and future professionals has become a critical priority, particularly in a context where technological evolution is outpacing existing educational structures.

In this context, this study examines how media professionals in the Basque Country perceive the opportunities and risks associated with AI, their current training levels, and what they consider necessary to adapt to this rapidly evolving landscape.

## 3. Materials and Methods

A mixed-method approach was employed, combining quantitative survey data with qualitative insights gathered through semi-structured in-depth interviews. This methodology was chosen to provide a comprehensive understanding of journalists' and media workers' experiences and perceptions while also quantifying trends and relationships across variables.

The survey was conducted online, with supplementary telephone support, during May and June 2024. A total of 504 responses were collected from journalists and communication sector professionals working in media, companies, or institutions in the Basque Country. Participants were identified through the Open Communication Guide provided by the Basque Government,1 which lists active media organisations in the region, and through the Basque Journalists' Association.

According to existing data (Basque Government, 2022; Pérez et al., 2023), it is estimated that approximately 5,000 people work in the media sector in Euskadi. Therefore, for a 95% confidence level, the margin of error of the survey is ±4.15%.

The survey employed an online panel and ensured respondent anonymity. No personal data were collected, eliminating risks related to data storage and handling. The sample included 276 men (55.1%), 223 women (44.5%), and two individuals who identified as "other" (0.4%). Variables such as years of professional experience, type of media, scope of the organisation, job role, and responsibility levels were also analysed.

In addition to the survey, six semi-structured in-depth interviews were conducted with senior figures specialising in technology, innovation, and AI from the most prominent media outlets based in the Basque Country. These interviews, lasting between 30 and 50 min, were conducted via videoconference in April and May 2024.

| Code | Media | Media Type | Position |
|---|---|---|---|
| I1 | Vocento / El Correo | Newspaper | Head of New Technological Trends and Quality |
| I2 | EiTB | Public broadcasting service | Deputy Director of EiTB Media |
| I3 | Noticias Group / Deia | Newspaper | Director of the Technology Department |
| I4 | Naiz | Newspaper | Content and Forecast Manager |
| I5 | Berria | Newspaper | Head of Information Technologies |
| I6 | Cadena SER / Radio Bilbao | Radio | Chief Editor |

Based on the state of the art, this study seeks to answer the following research questions:

RQ1: What specific AI training have journalists received so far?
RQ2: What are the perceived training needs among journalists in this area?
RQ3: What are the current policies of media organisations regarding AI training, according to media leaders?

## 4. Results

*4.1. AI training among media professionals*

The findings reveal that AI training levels among journalists are generally low. Despite the growing importance of this technology in the media industry, only 14.1% of professionals have received any form of training in this area. In terms of gender, women (15.9%) reported slightly higher training rates than men (12.7%).

Table 1. Professionals Who Have Received AI Training, by Media Type

|  | Men | | Women | | TOTAL | |
|---|---|---|---|---|---|---|
| **Type of media** | N | % | N | % | N | % |
| **Press** | 14 | 12,2 | 5 | 6,1 | 19 | 9,6 |
| **Radio** | 6 | 11,1 | 3 | 10,3 | 9 | 10,8 |
| **TV** | 2 | 5,7 | 2 | 5,9 | 4 | 5,8 |
| **Digital** | 6 | 17,6 | 11 | 34,4 | 17 | 25,8 |
| **Press Office** | 1 | 5,0 | 12 | 29,3 | 13 | 21,3 |
| **Advert. Agency** | 5 | 45,5 | 3 | 42,9 | 8 | 44,4 |
| **Broadcasting area** | N | % | N | % | N | % |
| **Local** | 8 | 22,9 | 2 | 4,3 | 10 | 12,2 |
| **Regional** | 21 | 10,9 | 23 | 17,7 | 44 | 13,6 |
| **National** | 6 | 16,2 | 6 | 15,8 | 12 | 16,0 |
| **International** | 0 | 0,0 | 5 | 41,7 | 5 | 21,7 |
| **TOTAL** | 35 | 12,7 | 36 | 15,9 | 71 | 14,1 |

Source: Own elaboration

Significant differences in AI training are observed across different types of media (Table 1). For instance, professionals engaged in roles further from traditional journalism, such as corporate communication and advertising, show a stronger commitment to AI training.

Among respondents, 44.4% of those working in advertising agencies and 21.2% of those in communication departments reported receiving AI training, far exceeding the levels observed among those in purely journalistic roles. A more business-oriented and economically driven approach to communication activities may explain this higher investment in AI training, as it offers opportunities to optimise both time and costs in content production.

A notable difference in AI training levels is also evident among professionals working in outlets that produce purely journalistic content. Those employed in digital-native sections or outlets reported higher training rates (25.8%) compared to their counterparts in more traditional media formats, such as radio (10.8%), print (9.6%), and television (5.8%).

In summary, the closer media professionals are to traditional journalistic tasks and outlets, the lower their levels of AI training tend to be. Conversely, training is significantly higher among individuals working in digital areas of media organisations. Similarly, the further removed their roles are from core journalistic activities and the closer they are to persuasive functions—such as advertising or corporate communication—the greater the likelihood of having received AI training.

The findings also highlight a correlation between the size and scope of media organisations and the AI training their employees receive (Table 1). Larger organisations with broader dissemination areas show higher training levels, with international outlets reporting nearly double the training rate (21.7%) of local outlets (12.2%). These results suggest that the adoption of AI in media may exacerbate the existing technological gap that has emerged since the onset of media digitalisation. Smaller outlets, in particular, are less prepared to meet this challenge.

The data further confirms that professionals more distanced from traditional journalistic tasks, such as editors and presenters, have received less AI training compared to those in technical roles or working in corporate communication and advertising (Table 2).

Table 2. Professionals Who Have Received AI Training, by Job Position

|  | Men | | Women | | TOTAL | |
|---|---|---|---|---|---|---|
| **Job type** | N | % | N | % | N | % |
| Editor | 22 | 13,5 | 21 | 14,1 | 43 | 13,8 |
| Presenter, host | 3 | 4,7 | 3 | 9,7 | 6 | 6,3 |
| Graphic (design, photographer) | 1 | 11,1 | 2 | 28,6 | 3 | 18,8 |
| Technical (cameraman, sound) | 2 | 100 | 0 | 0 | 2 | 66,7 |
| Producer, webmaster | 0 | 0 | 3 | 30,0 | 3 | 15,8 |
| Management | 3 | 23,1 | 2 | 28,6 | 5 | 25,0 |
| Public relations | 3 | 27,3 | 3 | 21,4 | 6 | 24,0 |
| Copywriter, creative | 0 | 0 | 1 | 25,0 | 1 | 20,0 |
| Other | 1 | 25,0 | 1 | 33,3 | 2 | 28,6 |
| **Experience** | N | % | N | % | N | % |
| Less than 5 years | 3 | 5,7 | 6 | 12,0 | 9 | 8,7 |
| From 5 to 10 years | 7 | 15,9 | 4 | 14,8 | 11 | 15,5 |
| From 10 to 20 years | 2 | 3,9 | 6 | 10,5 | 8 | 7,4 |
| More than 20 years | 23 | 18,0 | 20 | 21,7 | 43 | 19,5 |
| **Responsibility role** | N | % | N | % | N | % |
| Yes | 22 | 17,9 | 23 | 22,3 | 45 | 19,9 |
| No | 13 | 8,5 | 13 | 10,6 | 26 | 9,4 |
| TOTAL | 35 | 12,7 | 36 | 15,9 | 71 | 14,1 |

Source: Own elaboration

Moreover, it is unsurprising that individuals in leadership roles have, on average, received twice as much training as those without such responsibilities (19.9% compared to 9.4%). The disruptive nature of AI technology, its potential impact on work processes, and the availability of greater training opportunities are likely factors influencing this trend. This pattern suggests that more experienced professionals, particularly those in managerial or strategic positions, are more engaged in adopting and implementing AI technologies. Their roles often require staying updated on technological advancements to make informed strategic decisions.

Similarly, employees with longer tenures in their organisations are more likely to have received training. Their stability and seniority in the workplace, coupled with their presence in leadership roles, likely contribute to these higher training levels. Additionally, a perceived technological gap may have driven them to seek further training. These findings underline the importance of integrating AI training into the early and mid-stages of professional development to equip workers for technological challenges.

*4.2 AI training policies in media organisations*

AI training is a critical factor for adopting this technology effectively. Among respondents who reported never using AI, 43.3% cited a lack of training as the primary barrier.

The study found that only 8.5% of participants received AI training provided by their organisations, whether online or in person, compared to 12.9% who pursued training independently.

By organisation type, advertising agencies, digital media, and communication departments stand out for their commitment to AI training (Table 3). In contrast, traditional and journalistic media outlets have offered little training in this area to their employees.

Table 3. Type of AI Training Received, by Media Type

|  | Self online course | | Self face-to-face course | | Self-study | | Company online course | | Company face-to-face course | |
|---|---|---|---|---|---|---|---|---|---|---|
| **Type of media** | N | % | N | % | N | % | N | % | N | % |
| **Press** | 5 | 2,5 | 2 | 1,0 | 17 | 8,5 | 4 | 2,0 | 8 | 4,0 |
| **Radio** | 5 | 6,0 | 2 | 2,4 | 7 | 8,4 | 3 | 3,6 | 2 | 2,4 |
| **TV** | 2 | 2,9 | 0 | 0,0 | 4 | 5,8 | 0 | 0,0 | 3 | 4,3 |
| **Digital** | 13 | 19,7 | 5 | 7,6 | 14 | 21,2 | 8 | 12,1 | 8 | 12,1 |
| **Press Office** | 6 | 9,8 | 2 | 3,3 | 10 | 16,4 | 7 | 11,5 | 4 | 6,6 |
| **Advert. Agency** | 2 | 11,1 | 1 | 5,6 | 7 | 38,9 | 6 | 33,3 | 4 | 22,2 |
| **Broadcasting area** | N | % | N | % | N | % | N | % | N | % |
| **Local** | 4 | 4,9 | 3 | 3,7 | 8 | 9,8 | 3 | 3,7 | 5 | 6,1 |
| **Regional** | 21 | 6,5 | 6 | 1,9 | 37 | 11,4 | 15 | 4,6 | 16 | 4,9 |
| **National** | 7 | 9,3 | 2 | 2,7 | 11 | 14,7 | 8 | 10,7 | 5 | 6,7 |
| **International** | 2 | 8,7 | 1 | 4,3 | 4 | 17,4 | 2 | 8,7 | 3 | 13,0 |
| **TOTAL** | 34 | 6,7 | 12 | 2,4 | 28 | 11,9 | 29 | 5,6 | 60 | 5,8 |

Source: Own elaboration

This trend aligns with the types of roles that reported higher training levels (Table 4), which were predominantly technical and commercial positions—areas more peripheral to the production of current affairs content.

In terms of the scope of media operations, 13% of employees at international outlets reported seeking AI training independently, the highest percentage among all categories. This reflects a greater personal initiative to acquire AI skills in broader dissemination environments. However, despite the correlation between media reach and AI training, a significant proportion of professionals across all levels remain untrained. In local outlets, 87.8% of professionals reported no AI training, a figure that, while slightly lower, remains substantial in regional (86.4%), national (84%), and international (78.3%) contexts.

Technical professionals stand out as the most trained group in AI, with 66.7% having received training. This aligns with the technical nature of AI, which demands expertise in data analysis, programming, and advanced software use. In management and leadership roles, 25% of professionals reported AI training, highlighting its strategic importance for decision-making and innovation.

Media leaders, however, emphasise that an exclusively technical approach to AI is insufficient. They stress the need to integrate technical understanding with ethical considerations to address the core challenges of journalism. As one interviewee noted: "We need people in newsrooms trained in these topics, and, of course, at a technical level. But perhaps, also on an ethical level, it is important to have technical knowledge of AI [.] to know when and how to use it" (I4).

Following technical roles, graphic design professionals (18.75%) and production staff (15.8%) reported higher training levels. Advertising and corporate communication professionals also showed notable rates (16.7 and 14%, respectively). Journalists in editorial roles (13.7%) and presenters (6.3%) had the lowest training rates. According to media leaders, the limited immediate application of AI in fast-paced editorial work may explain this lower adoption. As one leader remarked: "Internally, it has not affected us yet; at the moment, we are not using anything. But given the breadth of what we cover, it could be useful, yes. Something has indeed been considered" (I6).

Table 5. Type of AI Training Received, by Job Position

|  | Self online course | | Self face-to-face course | | Self-study | | Company online course | | Company face-to-face course | |
|---|---|---|---|---|---|---|---|---|---|---|
| **Job type** | N | % | N | % | N | % | N | % | N | % |
| **Editor** | 21 | 6,7 | 7 | 2,2 | 38 | 12,1 | 16 | 5,1 | 17 | 5,4 |
| **Presenter, host** | 3 | 3,2 | 1 | 1,1 | 5 | 5,3 | 2 | 2,1 | 2 | 2,1 |
| **Graphic (design, photographer)** | 2 | 12,5 | 0 | 0,0 | 3 | 18,8 | 0 | 0,0 | 0 | 0,0 |
| **Technical (cameraman, sound)** | 0 | 0,0 | 0 | 0,0 | 2 | 66,7 | 0 | 0,0 | 1 | 33,3 |
| **Producer, webmaster** | 1 | 5,3 | 1 | 5,3 | 2 | 10,5 | 0 | 0,0 | 1 | 5,3 |
| **Management** | 2 | 10,0 | 1 | 5,0 | 4 | 20,0 | 3 | 15,0 | 3 | 15,0 |
| **Public relations** | 3 | 12,0 | 2 | 8,0 | 4 | 16,0 | 5 | 20,0 | 4 | 16,0 |
| **Copywriter, creative** | 0 | 0,0 | 0 | 0,0 | 1 | 16,7 | 1 | 16,7 | 1 | 16,7 |
| **Other** | 2 | 28,6 | 0 | 0,0 | 1 | 14,3 | 1 | 14,3 | 0 | 0,0 |
| **Experience** | N | % | N | % | N | % | N | % | N | % |
| **Less than 5 years** | 6 | 5,8 | 1 | 1,0 | 9 | 8,7 | 2 | 1,9 | 2 | 1,9 |
| **From 5 to 10 years** | 4 | 5,6 | 2 | 2,8 | 9 | 12,7 | 4 | 5,6 | 5 | 7,0 |
| **From 10 to 20 years** | 3 | 2,8 | 2 | 1,9 | 7 | 6,5 | 3 | 2,8 | 1 | 0,9 |
| **More than 20 years** | 21 | 9,5 | 7 | 3,2 | 35 | 15,8 | 19 | 8,6 | 21 | 9,5 |
| **Responsibility role** | N | % | N | % | N | % | N | % | N | % |
| **Yes** | 21 | 9,2 | 7 | 3,1 | 36 | 15,8 | 19 | 8,3 | 19 | 8,3 |
| **No** | 13 | 4,7 | 5 | 1,8 | 24 | 8,7 | 9 | 3,3 | 10 | 3,6 |
| **TOTAL** | 34 | 6,7 | 12 | 2,4 | 60 | 11,9 | 28 | 5,6 | 29 | 5,8 |

Source: Own elaboration

The analysis also reveals a strong overlap between individual initiative and organisational support for training. Among those who received any form of training, 52.1% did so through both personal and company efforts, 39.4% relied solely on self-directed learning, and only 8.5% received training exclusively from their employer. This suggests that corporate training initiatives largely benefit those already motivated to learn, leaving a majority without access to such opportunities. Traditional media outlets and more journalistic roles appear to have the most room for improvement in this regard.

This self-motivated approach aligns with comments from media leaders: "It does not all depend on companies; individuals need that drive to keep training and equip themselves with more tools to do their job. Continuous training is not just a company's responsibility; it is also part of the worker's responsibility" (I6).

The findings underscore the need for enhanced corporate training opportunities to provide media professionals with a deeper and more practical understanding of AI. Media leaders emphasise that this training must be ongoing: "After just 2 months, you are already outdated" (I1). Thus, training efforts cannot be one-off initiatives but must be sustained over time: "We need to keep updating our skills because technology evolves so rapidly. What was cutting-edge a year ago is likely obsolete today" (I2).

Given the rapid pace of change in AI, the focus has shifted to continuous professional development. As one leader noted: "The market demands far ahead of what educational institutions can offer. These are skills everyone working with AI should know. For example, you must be careful about what you provide to AI systems, as you could be sharing confidential internal company information" (I1).

While media leaders recognise the urgency of training to address AI-driven changes, they also stress the importance of foundational knowledge in technical aspects of AI (I2, I4, I5). One leader commented: "We need to improve competencies and prioritise training for technology use in general, as media organisations include teams of all ages with varying technological capacities" (I3).

## 5. Discussion and conclusions

This study on AI training among communication professionals in the Basque Country reveals a significant gap in specialised training, despite the growing importance of AI in the media industry (Husnain et al., 2024). The findings indicate that only a small percentage of journalists have received AI training so far, most of which was self-directed. The continuous training policies offered by companies and the self-directed learning initiatives of journalists emerge as essential needs in an environment undergoing rapid and profound transformation.

The disparities identified among different types of organisations and professional profiles highlight key areas where training efforts should be directed. Larger communication companies with broader dissemination reach have made greater investments in training their professionals, while smaller, local organisations have so far made fewer efforts in this area. This disparity suggests that AI could further widen the technological gap already created by media digitalisation, intensifying the divide between major international platforms and smaller, resource-constrained outlets.

Similarly, the data also reveal a differential adoption of AI between core and peripheral aspects of media operations. While areas adjacent to journalistic activity (technical, commercial, etc.) are adopting this technology earlier through training, professionals directly involved in content creation are lagging behind in terms of training received. This may be influenced by scepticism among journalists about the potential impact of AI on their profession (Peña-Fernández et al., 2023b).

Media leaders in the Basque Country acknowledge the need for greater ongoing training in AI (Noain Sánchez, 2022). Technological training is viewed as essential, not only to enhance individual skills but also to ensure the responsible and effective use of AI in media. In the short term, organisations are advised to provide in-house or company-supported training programmes to enable collaboration between media professionals and technical staff, fostering awareness of the risks associated with improper AI use. In the longer term, university-level training programmes should also address the broader implications of AI applications in media, focusing on its potential to generate misinformation, perpetuate biases, and exacerbate social inequalities (Opdahl et al., 2023).

In summary, this research highlights the urgent need to expand AI literacy and training among communication professionals (Deuze and Beckett, 2022; Gómez-Diago, 2022; Larrondo-Ureta and Peña-Fernández, 2024). Only through continuous and comprehensive AI training—encompassing both technical and ethical aspects—can media professionals effectively meet contemporary challenges and ensure the responsible use of AI in the media industry.